\documentclass[11pt]{article}

\usepackage{a4wide}
\usepackage{braket}
\usepackage{amsmath,amssymb}
\usepackage{enumerate}
\usepackage{graphicx,epstopdf}
\usepackage{color}

\DeclareMathOperator*{\Res}{Res}
\DeclareMathOperator{\im}{Im}
\DeclareMathOperator{\re}{Re}
\newcommand{\nn}{\ensuremath{\nonumber\\}}

\newcommand{\be}{\begin{equation}}
\newcommand{\ee}{\end{equation}}
\newcommand{\bea}{\begin{eqnarray}}
\newcommand{\eea}{\end{eqnarray}}
\newcommand{\ba}{\begin{array}}
\newcommand{\ea}{\end{array}}

\newcommand{\alp}{{\alpha'}}

\newcommand{\cA}{{\cal{A}}}

\newcommand{\cP}{{\cal{P}}}

\begin{document}

\title{
	Geometric cross sections of \\
	rotating strings and black holes \bigskip}

\author{
        Toshihiro Matsuo$^*$ and
	Kin-ya Oda$^\dagger$\bigskip\\
	$^*$\footnotesize\it Graduate School of Pure and Applied Sciences, University of               Tsukuba, Ibaraki 305-8571, Japan\\
	\footnotesize\tt tmatsuo@het.ph.tsukuba.ac.jp\smallskip\\
	$^\dagger$\footnotesize\it Department of Physics, Osaka University, Osaka 560-0043, Japan\\
	\footnotesize\tt odakin@phys.sci.osaka-u.ac.jp\bigskip}

\date{September 24, 2008}

\maketitle

\bigskip

\begin{abstract}
\noindent\normalsize
We study the production cross section of a highly excited string with fixed angular momentum from an ultra-high energy collision of two light strings.
We find that the cross section exhibits geometric behavior in a certain region of angular-momentum/impact-parameter space.
This geometric behavior is common to the differential cross sections of a black hole production with fixed angular momentum and thus we see another correspondence between strings and black holes.
\end{abstract}

\vfill

{\hfill OU-HET 610/2008}

\newpage

\section{Introduction}

String theory not only cures the ultraviolet divergences from graviton loops in S-matrices but also has revealed nontrivial consequences of black hole physics in which both gravity and quantum mechanics play pivotal roles together, adding credibility as a candidate for the theory of quantum gravity.
Especially it is claimed that there is a correspondence: a black hole should correspond to an ensemble of highly excited single string states at a certain threshold $g_s^2M\sim1$, namely the correspondence point, with $g_s$ and $M$ being the (closed) string coupling constant and the mass, respectively.\footnote{In this paper, we shall work in string natural unit $\alp=1,4$ etc. We will specify when necessary.}
The string and black hole pictures should be valid for $g_s^2M\ll1$ and $g_s^2M\gg1$, respectively.
This is based on the observation that both the string and black hole descriptions give the same order of entropy at the correspondence point, as one varies adiabatically the string coupling constant $g_s$ for a fixed mass $M$~\cite{Susskind:1993ws,Horowitz:1996nw}.
When one takes into account the self interactions, the string gets entangled in itself and consequently its typical size is reduced to the order of the string length scale which coincides with the horizon radius of the black hole at the threshold, providing another support for the correspondence~\cite{Horowitz:1997jc,Damour:1999aw}.

It should be interesting to consider the correspondence between a formation process of a string and that of a black hole both from an ultra high energy scattering, since the latter involves non-perturbative dynamics of the gravitational interactions which is not successfully formulated within string theory so far.\footnote{We also note that in a scenario with the string scale around TeV~\cite{Antoniadis:1998ig,Randall:1999ee}, this process is not only theoretically important but also directly testable at the CERN Large Hadron Collider and beyond~\cite{Giddings:2001bu,Dimopoulos:2001hw}. See also~\cite{Oda:2001uw}.}
At low energies, the string picture should be valid while at high energies, the black hole picture would prevail.
For a classical gravitational scattering with center of mass energy $\sqrt{s}$, the black hole production cross section has been proven to be of the order of the black disk with its size being the Schwarz\-shild radius $r_S\sim(g_s^2M)^{1/(D-3)}$ of the black hole mass $M\sim\sqrt{s}$, where $D=d+1$ is the number of (large) spacetime dimensions~\cite{Eardley:2002re,Yoshino:2002tx,Yoshino:2005hi}.
Physical interpretation is that when initial particles are wrapped within the horizon scale, a black hole forms~\cite{'t Hooft:1987rb}.

Dimopoulos and Emparan~\cite{Dimopoulos:2001qe} have investigated the production of a single highly excited string as a black hole progenitor, in view of the correspondence principle with a fixed string coupling $g_s\ll1$ and with a varying mass $M$, or equivalently with a varying center of mass energy $\sqrt{s}\sim M$. At the tree level, they obtained a linearly raising total cross section with respect to $s$ (for closed string). 
The tree level cross section does not match the black hole one at the correspondence point $\sqrt{s}\sim M_C\equiv g_s^{-2}$.
Actually the resultant string cross section hits the unitarity bound around $\sqrt{s}\sim g_s^{-1}$ which is below $M_C$.
Historically, when string theory was applied to the strong interactions, it was conjectured, based on  experimental observations, that the theory would provide a constant total cross section above the unitarity bound if one managed to take all the higher order loop corrections into account~\cite{Nambu-san}.
If it is the case, the constant cross section, that is to be of order one, matches the black hole one at the correspondence point.
Currently one cannot prove the validity of this argument given the status of string theory where the cross section is computed only up to few loops and the relative phase of each loop cannot be fixed a priori. 
Also, the total cross section of open string turns out to be constant of order $g_s$ at tree level, as we will see below, and again it does not agree with that of the black hole at the correspondence point.\footnote{The correspondence principle is blind to the end points of strings. 
There does not seem to be a strong reason to exclude the gravitating string ball made of a long open string when self interactions including closed string exchanges are taken into account at higher orders.
 }

However, it is just a first step to study a total cross section that is a sum of infinitely many partial wave cross sections.  
By decomposing a total cross section into partial wave cross sections we are able to investigate the correspondence in more detail.
When a black hole is produced from a high energy scattering, two initial particles have a finite impact parameter~$b$ and therefore finite angular momentum
$J\sim b|\vec{p}|\sim bM$. 
If one assumes that the initial angular momentum is not lost much during the black hole formation process, the differential cross section $d\sigma_\text{BH}/dJ$ increases geometrically with the angular momentum as $J^{D-3}$ until a threshold value $J_\text{max}$~\cite{Ida:2002ez}.
On the other hand, a string partial wave cross section is considered to give exponential damping with respect to $J$ due to the softness of string at high energy.
Thus at first sight there does not seem a correspondence to hold even at the partial wave level.

In this paper, we revisit the partial wave string amplitude in the ultrahigh energy limit, and show that the softness of string is seen at relatively large angular momentum or impact parameter\footnote{See also Refs.~\cite{Veneziano:2004er,Veneziano:2006wn} for related arguments on string side with a finite impact parameter.} while in a certain region of the angular momentum space the partial wave cross section indeed shows the geometric behavior for both closed and open strings thus we find an universal behavior that is comparable with that of black hole.

The organization of the paper is as follows:
In section~2, we present our computation of the production cross section of an excited string with fixed total angular momentum $J$.
We find soft behavior characteristic to a string in the Regge limit at large angular momentum region as well as a geometric behavior of the partial wave cross section for an angular momentum which is rather large but less than a certain energy scale in string unit. 
Then we discuss the energy region where the partial wave unitarity condition is satisfied, which is the necessary condition for the perturbative expansion to be valid.
We also argue open string case. 
In section~3, we briefly review the correspondence principle and discuss the production cross section of a rotating black hole. Then we apply the correspondence principle to the black hole/heavy string production processes.
Section~4 provides the summary and discussions for possible future directions.
In Appendix~\ref{sigma_from_residue}, we review an alternative way to obtain cross sections through residue computation. 
In Appendix~\ref{Gegenbauer_formulas}, we collect formulae useful for our computations.
In Appendix~\ref{partial_wave_unitarity_bound}, we review the derivation of the partial wave unitarity condition in $D$ dimensions.

\section{Rotating string production}

In order to get the production cross section of a string with fixed total angular momentum~$J$, we shall employ four tachyon tree amplitude at high energy.
For closed strings, the imaginary part of the amplitude provides the fixed angle total cross section whereas the real part dominantly governs the unitarity condition.

\subsection{Total cross section}
We consider the four tachyon tree amplitude in closed bosonic string theory. 
The amplitude is given by the Virasoro-Shapiro amplitude:
\begin{align}
	\cA^\text{closed}(s,t)	&=	
	2\pi g_s^2{\Gamma(-1-s)\Gamma(-1-t)\Gamma(-1-u)\over\Gamma(2+s)\Gamma(2+t)\Gamma(2+u)},
		\label{Virasoro_Shapiro}
  \end{align}
where $s+t+u=-4$ with $\alpha'=4$.

The high energy process in our concern is controlled by the Regge limit $s \gg 1$ with a fixed $t$.
The amplitude in the Regge limit is computed by Stirling's formula $\Gamma(n+1)\simeq n^ne^{-n}\sqrt{2\pi n}$ and become\footnote{\label{Physical_sheet} This is obtained by the prescription which simulates the expected quantum corrections to the sharp tree level resonances on the physical sheet (along real axis in the complex $s$ plane). The Regge limit is taken on the second sheet avoiding the poles on the real axis. See also footnote~\ref{off_axis}. }
\begin{align}
  \cA^\text{closed}(s,t)
  	&\to	2\pi g_s^2\,s^{2+2t}\left(-{1\over t}+i\pi\right)
			\quad \text{for $s\gg1$} .
\label{Regge_limit}
\end{align}

Once the asymptotic form of high energy amplitude is found it is straightforward to see the total cross section for production of a heavy string state.
The optical theorem provides the total cross section from the imaginary part of the forward scattering:\footnote{In \cite{Dimopoulos:2001qe} the authors obtained \eqref{total_cross_section} by computing the residues of the $s$-channel resonances and averaging the delta function (see also Appendix~\ref{sigma_from_residue})
\begin{align*}
\sigma^\text{closed}(s) \simeq -\frac{\pi}{s} \Res \cA^\text{closed}(s,t=0) = 2\pi^2 g_s^2 s .
\end{align*}
}
\bea
\sigma^\text{closed}(s) = \frac{1}{s} \im \cA^\text{closed}(s,t=0) = 2\pi^2 g_s^2 s .
\label{total_cross_section}
\eea
The total cross section raises linearly with $s$, as opposed to the field theory cases in which total cross sections decrease with energy due to the uncertainty relation: the higher the energy of particle, the smaller its wave length, i.e.\ the smaller effective size of the scatterer. 
On the other hand, the stringy uncertainty relation~\cite{Yoneya:2000bt} indicates that the higher the scattering energy, the bigger the area of string becomes, resulting in the growing cross section. 
Intuitively, one may think that the raising cross section implies a breakdown of unitarity. However string theory (and gravity) contains massless modes and thus has a long range potential. 
Therefore there is no known restriction for total cross sections with respect to the unitarity such as the Froissart bound, thus the linearly raising behavior~\eqref{total_cross_section} is not necessary a contradiction to any physical requirement.\footnote{
In \cite{Gross:1987kza, Gross:1987ar}, the genus $G$ four point amplitude in the Regge limit is computed as
\begin{align*}
{\cal A}_G(s,t) \simeq g_s^{2G+2} (\ln s)^{-12 G} s^{G+2} , 
\end{align*}
where the $t$ dependence is not determined and more importantly the relative phase is not also known. 
One might suppose the relative phase of the leading correction to the tree amplitude is pure imaginary and then
\begin{align*}
\sigma(s) \simeq g_s^2 s - g_s^4  (\ln s)^{-12} s^2 +O(g_s^6) ,
\end{align*}
which seems to satisfy the unitarity condition at high energy.
Although it would be interesting to consider such higher order corrections, we do not pursue them and shall push forward the tree level analysis in this paper.}
To argue the unitarity we need to decompose into partial waves which we shall investigate in the following.

\subsection{Partial wave cross section---rotating string production}
Let us compute the production cross section of a heavy string state with a fixed total angular momentum $J$. 
To this end we shall consider the partial wave expansion of the amplitude.
In $D$ dimensions, the spherical harmonics is given by the Gegenbauer polynomials and the partial wave expansion of the amplitude is given in terms of them
\begin{align}
	\cA(s,t)	&=	\sum_{J=0}^{\infty}\cA_J(s)
				{C_J^{\nu}(\cos\theta)\over C_J^\nu(1)},\label{partial_wave_expansion}
  \end{align}
where $\nu=(D-3)/2$ and $\cos\theta=1+2t/s$ with $\theta$ being the scattering angle. 
In~\eqref{partial_wave_expansion}, we put the factor
$C_J^\nu(1)={\left(2\nu\right)_J/\Gamma(J+1)}$
to yield $\cA(s,0)=\sum_{J=0}^\infty\cA_J(s)$ so that the simple normalization
\begin{align}
\sigma_J(s)={1\over s}\im\cA_J(s)
\end{align}
leads to the required formula $\sigma(s)=\sum_{J=0}^\infty\sigma_{J}(s)$, where $\left(a\right)_n=\Gamma(a+n)/\Gamma(a)$ is the Pochhammer symbol.
Using the orthogonality condition~\eqref{Gegenbauer_orthogonality}
with the normalization factor~\eqref{normalization_factor} in Appendix \ref{partial_wave_unitarity_bound},
the partial wave amplitude is given by\footnote{
	When $D=4$, the expansion reduces to the better-known formula with the Legendre polynomial
	\begin{align*}
		\cA(s,t)	&=	\sum_{J=0}^{\infty}\cA_J(s)\,P_J\left(1+2t/s\right), &
		\cA_J(s)
				&=	{2J+1\over2}
					\int_{-1}^1 d\cos\theta\,P_J(\cos\theta)\,
						\cA\left(s,-s{1-\cos\theta\over2}\right).
	  \end{align*}}
\begin{align}
	\cA_J(s)
		&=	{C_J^\nu(1)\over
				N_J^\nu}
			\int_{-1}^1 dz\left(1-z^2\right)^{\nu-{1\over2}}C_J^\nu(z)\,\cA(s,t),
			\label{partial_wave_amplitude}
  \end{align}
where $t=-s(1-z)/2$ (and $z=\cos\theta$).

In the high energy limit $s\gg1$, the integral~\eqref{partial_wave_amplitude} is dominated by the forward scattering region $|t|/s\ll1$ or $1-z\ll1$.
Thus the Regge limit \eqref{Regge_limit} yield good approximations for the integral~\eqref{partial_wave_amplitude} and we have: 
\begin{align}
	\cA_J^\text{closed}(s)
		&\simeq	2\pi g_s^2\,s^2{C_J^\nu(1)\over
				N_J^\nu}
			\int_{-1}^1 dz\left(1-z^2\right)^{\nu-{1\over2}}C_J^\nu(z)\left(-{1\over t}+i\pi\right)s^{2t} .
\end{align}
The production cross section of a highly excited string with total angular momentum $J$ is given by the partial wave cross section through the optical theorem
\begin{align}
\sigma_J^\text{closed}(s)=
{1\over s}\im\cA_J^\text{closed}(s)
=2^{\nu+1}\pi^{5 \over 2} g_s^2 \Gamma\!\left(\nu+{1\over2}\right)
{\left[C_J^\nu(1)\right]^2\over N_J^\nu}
{s\,e^{-\lambda} \over \lambda^\nu}I_{J+\nu}(\lambda),
\label{sigmaJbeforeapp}
\end{align}			
where $\lambda\equiv s\ln s$ and we have utilized Eq.~\eqref{Gegenbauer_integral_plus} in Appendix \ref{Gegenbauer_formulas}  to perform the integration.
Further from the limit~\eqref{I_limit} in $\lambda\gg1$ for the modified Bessel function, we get\footnote{ 
In~\cite{Kuroki:2007aj} similar expression is obtained. However the weight function that appears in the partial wave expansion is not specified there, thus they obtained the partial wave cross section up to the $J$-dependent coefficient which is fixed in the present paper. Actually this factor is the origin of  the geometric behavior of the cross section.}
\begin{align}
\sigma_J^\text{closed}(s)
&\simeq 2^{\nu+{1\over 2}}\pi^2 g_s^2  \Gamma\!\left(\nu+{1\over2}\right)
{\left[C_J^\nu(1)\right]^2\over N_J^\nu}{s \over \lambda^{\nu+{1\over2}}}{e^{-{(J+\nu)^2\over2\lambda}}} .
\label{sigmaJ}
\end{align}

The exponential factor shows the softness of the cross section that is a characteristic feature of string in the high energy processes. 
We can introduce an impact parameter $b$ through the total angular momentum with a fixed center of mass energy as $b=J/\sqrt{s}$.
We find the effective size of string is about $\sqrt{\ln s}$. 
This is consistent with the well-known argument that the Fourier transform of the Regge amplitude with respect to the transverse momentum $p_{\perp}^{2} \sim -t$ gives the effective size of the string in the transverse space resulting the gaussian profile of width $\sqrt{\ln s}$:
\begin{align}
{1\over s}\int {d^{D-2} p_{\perp}\over (2\pi)^{D-2}} A_\text{Regge} e^{i p_{\perp} \cdot x} 
&= s(4\pi \ln s)^{-(D-2)/2} e^{-x^{2}/(4\ln s)} ,
&
A_\text{Regge}
&\sim s^{2+2t} .
\label{Fourier_transform}
\end{align}
Note that this argument is only valid at large $x$ because $t$ is small in the Regge amplitude and that it is not reliable at small $x$ region.
On the other hand, Eq.~\eqref{sigmaJ} is valid for all $J$ and thus we are able to investigate the small $b$ ($=J/\sqrt{s}$) region.
Therefore it is interesting to see the behavior in Eq.~\eqref{sigmaJ} especially at $J \lesssim \sqrt{s\ln s}$ where the gaussian damping factor can be neglected.
Recalling the normalization~\eqref{normalization_factor} in Appendix \ref{Gegenbauer_formulas} and $C_J^\nu(1)=(2\nu)_J/\Gamma(J+1)$, one may immediately read from Stirling's formula for large $J \gg 1$ but $J \lesssim \sqrt{s \ln s}$ that the cross section behaves geometrically
\bea 
\sigma_{J}(s) \propto J^{D-3}.
	\label{sigmaJ_app}
\eea
Thus we conclude that at $1 \ll b \lesssim \sqrt{s}$ the cross section is described by a black disc.
As we will argue, this is the characteristic behavior of differential cross sections of black hole and thus we have a correspondence.

Before proceeding, several comments are in order:
\begin{itemize}
\item
So far we have considered the partial amplitude decomposed by using a basis of the Gegenbauer polynomials that is the highest spherical harmonics in $D$ dimensions. This means that we have obtained partial wave amplitudes with respect to the total angular momentum $J$.
However we may use the ``lowest spherical harmonics,'' namely a plane wave $e^{i J_{12}\theta}$, which is an eigenfunction of the angular momentum $J_{12}$.
In this case one has
\begin{equation}
\sigma_{J_{12}}	=	{1\over s}\int_0^{2\pi}{d\theta \over 2\pi} \im A(s,t) e^{-iJ_{12}\theta},
\qquad 
\text{with}\quad t=-{s\over2}(1-\cos\theta).
\end{equation}
In other words, the $J_{12}$ partial wave cross section is obtained from the Fourier transform of the imaginary part of the amplitude \cite{Kuroki:2007aj}.
Although this is the same form as Eq.~\eqref{Fourier_transform} with $D=3$,  these are different Fourier transformations.             
As before, we may introduce an impact parameter (projected onto 1-2 plane): $b_{12}=J_{12}/\sqrt{s}$ and may focus on the small~$\theta$ region (as a consequence of the Regge amplitude $|t| \ll 1$) by defining $\tilde{\theta}\equiv\sqrt{s}\theta$:
\begin{align}
\sigma_{J_{12}}
	&\sim	{1\over s}\int_0^{2\pi\sqrt{s}}{d\tilde{\theta} \over 2\pi} e^{-{\tilde{\theta}^2 \over 2}\ln s -ib_{12}\tilde{\theta}}
	=		\sqrt{s\over 2\pi \ln s}e^{-{b_{12}^2\over 2\ln s}} .
\end{align}

\item 
The cross section~\eqref{sigmaJ} has been obtained by applying the Regge limit to the integrand. 
On the other hand, as we explain in Appendix~\ref{sigma_from_residue}, one can estimate its production cross section with fixed angular momentum $J$ by reading off the residue
\begin{align}
	\sigma_{J}(s)
		&\simeq
			-{\pi\over s}\Res_{s=N} \cA_J(s)\nonumber\\
		&=
			{2 \pi^2 g_s^2 \over s}
			{C_J^\nu(1)\over N_J^\nu}
			\int_{-1}^1 dz\,(1-z^2)^{\nu-{1\over2}}C_J^\nu(z)\left(\Gamma(s+t+3)\over\Gamma(s+2)\Gamma(t+2)\right)^2.
			\label{partial_wave_cross_section}
  \end{align}
Note that the Regge limit has not been taken here. 
The integral can be evaluated numerically and we can compare \eqref{sigmaJbeforeapp} and \eqref{partial_wave_cross_section} to check the validity of the  Regge limit in the integrand for a given $s$. 
As an illustration, the result for $s=100$ is shown in Fig.~\ref{sigma_fig}.
\begin{figure}
	\begin{center}
		\mbox{\includegraphics[width=.5\linewidth]{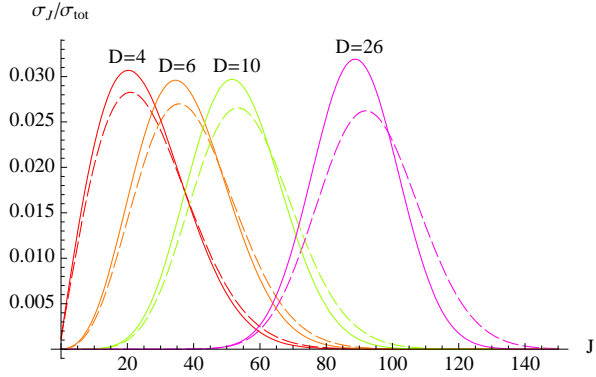}}
		\caption{Cross section $\sigma_J/\sigma_\text{tot}$ for $s=100$. $\sigma_J$ is obtained by Regge limit~\eqref{sigmaJbeforeapp} and by reading off the residues~\eqref{partial_wave_cross_section} (dashed and solid lines, respectively). $\sigma_\text{tot}$ is given by~\eqref{total_cross_section}.}
		\label{sigma_fig}
	  \end{center}
  \end{figure}

\end{itemize}

\subsection{Unitarity bound}
For completeness, we discuss how large $s$ the cross section can be trusted via the unitarity argument.
We will study the partial wave unitarity by analyzing the four point amplitude from which we read off the production cross section of a string with fixed angular momentum.
The partial wave unitarity at tree level has been investigated in~\cite{Soldate:1986mk} (and beyond tree level with eikonal approximation in~\cite{Muzinich:1987in}), see also~\cite{Nambu:1969sb}. It has been shown that when the angular momentum is less than $\sqrt{s \ln s}$, the partial wave amplitude might break the unitarity bound at high energy.
Here we check whether our geometric behavior of the amplitude is within the  unitarity bound.

Due to the $t$-channel exchange of the massless ($t=0$) graviton, we see from Eq.~\eqref{Regge_limit} that the real part will be larger than the imaginary part in Eq.~\eqref{partial_wave_amplitude} in the high energy limit $s\gg1$.
In the limit, the real part can be written as
\begin{align}
	\re\cA_J^\text{closed}(s)
		&=	4\pi g_s^2{C_J^\nu(1)\over N_J^\nu}s\,e^{-s\ln s}\int_0^\infty dw\,e^{-w}\int_{-1}^1dz\left(1-z^2\right)^{\nu-{1\over2}}C_J^\nu(z)e^{(s\ln s+w)z}.
  \end{align}
Using the formula~(\ref{Gegenbauer_integral_plus}) in Appendix \ref{Gegenbauer_formulas}, we get
\begin{align}
	\re\cA_J^\text{closed}(s)
		&=	2^{\nu+2}\pi^{3\over 2}g_s^2\,\Gamma\!\left(\nu+{1\over2}\right){{C_J^\nu(1)}^2\over N_J^\nu}s\,e^{-s\ln s}\int_0^\infty dw\,e^{-w}
			{I_{J+\nu}(s\ln s+w)\over\left(s\ln s+w\right)^{\nu}}.
  \end{align}
The modified Bessel function takes the limit $I_{J+\nu}(\lambda)\to e^{\lambda}/\sqrt{2\pi\lambda}$ for $\lambda\gg1$ and we get
\begin{align}
	\re\cA_J^\text{closed}(s)
		&=	2^{\nu+{3\over2}}\pi g_s^2\,\Gamma\!\left(\nu-{1\over2}\right){{C_J^\nu(1)}^2\over N_J^\nu}{s\over(s\ln s)^{\nu-{1\over2}}}.
			\label{real_part_of_A}
  \end{align}
With the normalization of Eq.~\eqref{normalized_amplitude}, we have
\bea
	\re a_J^\text{closed}(s)
={g_s^2s\over \left(2\nu-1\right) 
2^{3\nu+{1\over2}} \pi^{\nu-{1\over2}}
\left(\ln s\right)^{\nu-{1\over2}}
} .
\label{real_part_of_normalized_amplitude}
\eea
On the other hand the imaginary part is also written with the normalization of Eq.~\eqref{normalized_amplitude} as
\bea
\im a_J^\text{closed}(s) 
=
{g_s^2 s \over 2^{3\nu+{5\over 2}} \pi^{\nu-{3\over 2} } \left(\ln s\right)^{\nu+{1\over 2}}
} .
\label{imaginary_part_of_normalized_amplitude}
\eea
Plugging these into the condition for the partial wave unitarity~\eqref{unitarity_condition_for_normalized_amplitude}, we have 
\begin{equation}
{g_s^2s\over \left(\nu-{1\over 2}\right)^2 
2^{3\nu+{1\over2}} \pi^{\nu+{1\over2}}
\left(\ln s\right)^{\nu-{3\over2}}} \lesssim 1 \qquad 
\text{for}\quad \nu >{1\over 2} \quad (D >4),
\label{unitarity_condition_for_closed}
\end{equation}
where we have used the fact that the real part dominates over the imaginary part in the right hand side of~\eqref{unitarity_condition_for_normalized_amplitude} in Appendix \ref{partial_wave_unitarity_bound}. 
Indeed the real part is larger than the imaginary part by a factor $\ln s$.
This shows that no matter how small the string coupling is, the unitarity bound will be violated at sufficiently high energy.

It is interesting to notice that up to the $\ln s$ factor the partial wave unitarity bound is hit at $s \simeq 1/g_{s}^{2}$ which is precisely at the total cross section \eqref{total_cross_section} becomes of order one.
When one includes the $\ln s$ factor, the total cross section is reliable up to larger value of $s$ for a fixed $g_s$.

\subsection{Open string case}
So far we have considered the closed string scattering.
We can repeat the above argument for the open string case.
We start from the Veneziano amplitude:
\begin{equation}
	\cA^\text{open}(s,t)	=	g_s \left[
	{\Gamma(-\alpha(s))\Gamma(-\alpha(t))\over\Gamma(-\alpha(s)-\alpha(t))}
	+{\Gamma(-\alpha(t))\Gamma(-\alpha(u))\over\Gamma(-\alpha(t)-\alpha(u))}
	+{\Gamma(-\alpha(u))\Gamma(-\alpha(s))\over\Gamma(-\alpha(u)-\alpha(s))}\right],
		\label{Veneziano}
  \end{equation}
where $\alpha(x)=1+\alp x$ and $s+t+u=-4$ with $\alpha'=1$.
The Regge limit reads\footnote{\label{off_axis} We take the large $s$ limit off the real axis $s\to(1+i\epsilon)\infty$ so that $\left(\sin\pi\alpha(s)\right)^{-1}\to0$ and $\left(\tan\pi\alpha(s)\right)^{-1}\to-i$ exponentially. Physically, this limit corresponds to the assumption that the $s$-channel resonances have decay widths that increase at least linearly with the pole masses, which is natural given the exponentially growing number of decay modes in string theory. See also footnote~\ref{Physical_sheet}.}
\begin{align}
	\cA^\text{open}
		&\to	{-g_s \pi \over \Gamma\left(1+\alpha(t)\right)\sin\pi \alpha(t)}\left(1+e^{-i \pi \alpha(t)}\right) \left(\alpha(s)\right)^{\alpha(t)} & (s\gg1)
\nonumber \\
		&\simeq \pi g_s s\left({\pi \over 2} t + i \right) & (t \simeq 0),
\end{align}
where the last line is the case in which the amplitude is dominated over by the $t$-channel exchange of massless modes of the open string.
Note that the real part is small compared with the imaginary part contrary to the closed string case.

We find immediately the total cross section:
\bea
\sigma^\text{open}(s) = \frac{1}{s} \im \cA^\text{open}(s,t=0) = \pi g_s .
\eea 
The partial wave cross section is obtained quite similarly as in the case of closed string.
The imaginary part gives
\bea
\im\cA_J^\text{open}(s)
=
2^\nu \pi^{{3 \over 2}} \Gamma\!\left(\nu+{1\over2}\right) g_s {\left[C_J^\nu(1)\right]^2\over N_J^\nu} 
{s \,e^{-\lambda'} \over \lambda'^\nu}I_{J+\nu}(\lambda'),
\label{Im_A_open}			
\eea
where $\lambda'=(s \ln s) /2$.
Again we find a geometric cross section at $1 \ll J \leq \sqrt{s\ln s}$ 
\bea
\sigma_J^\text{open} \simeq J^{D-3}.
\label{sigmaJ_app_open}
\eea

We may also check the partial wave unitarity.
The real part is negligible to the imaginary part and thus the partial wave unitarity condition $\im a_J \geq |a_J|^2$ becomes 
\bea
\im a_J^\text{open} \leq 1.
\eea
From Eq. (\ref{Im_A_open}) we have
\bea
\im a_J^\text{open}= {g_s \over 2^{2\nu+3} \pi^{\nu+{1\over 2}} (\ln s)^{\nu+{1\over2}}} 
\eea
thus the partial wave unitarity condition is satisfied at any high energy for open strings.

\section{Correspondence and black hole cross section}

Let us briefly review the correspondence principle for black hole and string~\cite{Susskind:1993ws, Horowitz:1996nw}.
In the perturbative formulation of string theory, the string coupling constant $g_s$ is a free parameter, which should be fixed by a dilaton vacuum expectation value supposedly fixed by non-perturbative effects.
When we vary $g_s$ and/or the mass $M$, it is observed that various physical quantities are smoothly transited between string and black hole pictures.
Especially as we vary the coupling adiabatically, adiabatic invariants would not change drastically at the transition point.
The order parameter for the transition is $\mu\equiv g_s^2M$.
The larger the coupling and/or mass is, the stronger the gravitational interactions are.
Therefore we will see that the black hole (string) picture becomes valid for $\mu\gg1$ ($\ll1$) and the transition point is given at $\mu \sim 1$.
The gravitational constant is given by $G\sim g_s^2$ in any dimensions.
When $D=d+1$ spacetime dimensions are large, that is, uncompactified or compactified with a length scale much larger than our region of interactions, the $D$-dimensional Planck mass (length) is given by $M_D\sim g_s^{-2/(D-2)}$ ($\sim \ell_D^{-1}$).
Note that, for a string coupling being fixed at a small value $g_{s}<1$, the Planck scale is always smaller than the correspondence scale $M_D\lesssim M_C=g_s^{-2}$ and that the black hole picture should be valid at the correspondence point $M\sim M_C$.

The Schwarzschild radius for a black hole with mass~$M$ is given by $r_S\sim(GM)^{1/(D-3)}\sim\mu^{1/(D-3)}$ and its entropy is given by the horizon area $S_\text{BH}\sim r_S^{D-2}/G\sim (g_s^2M^{D-2})^{1/(D-3)}$.
For free string states with a fixed mass~$M$ (being equal with its length in string unit), the entropy becomes $S_\text{string}\sim M$.
Therefore the entropy becomes the same order in both pictures at the correspondence point $\mu\sim 1$ for any number of (large) dimensions.\footnote{We note that the self interactions of the string are neglected to obtain $S_\text{string}\sim M$. At this level, given the form $S_\text{BH}\propto M^a$ with $a$ being some constant, the agreement of the temperature $T=(\partial S/\partial M)^{-1}$ at the correspondence point is trivially derived from that of the entropy. See also  \cite{Lin:2007gi}.}
We note that the string states are treated within a micro/grand canonical ensemble and that a black hole corresponds to the ensemble for the correspondence to hold. The compared quantities are averaged values within the ensemble.
If we neglect the self interactions of the string, typical size would be that of the random walk $R_\text{RW}\sim \sqrt{M}$, which is much larger than the black hole horizon radius at the correspondence point for $g_s<1$.
This discrepancy can be solved by properly taking into account the self interactions~\cite{Horowitz:1997jc,Damour:1999aw}.

If correspondence is valid and a black hole can be viewed as an ensemble of the corresponding string states, black hole production cross sections must be connected to the production cross sections of string states.
As we have seen in section two, the tree level total production cross section of a closed string is obtained from the four point amplitude and the result is $\sigma_\text{tot}(s)\sim g_{s}^{2}s$. 
When $s\gtrsim g_{s}^{-2}$ the partial wave cross sections start to exceed the unitarity bound~\eqref{unitarity_condition_for_closed}.
We note that this is the scale where the D-brane interactions become also significant.

\begin{figure}
	\begin{center}
		\mbox{\includegraphics[width=.5\linewidth]{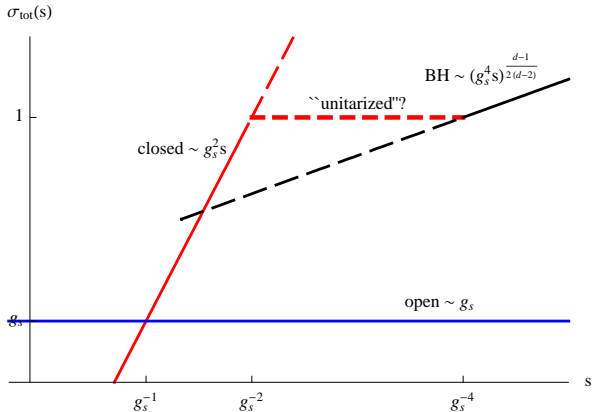}}
		\caption{Schematic log-log plot for the production cross section of a string and black hole.}
		\label{correspondence_fig}
	  \end{center}
  \end{figure}

Now let us consider the trans-Planckian collision of two light particles.
The production cross section of a black hole can be computed by the assumption that the colliding target looks like a black disk with its radius being of order the Schwarzschild radius $r_S\sim(g_s^2M)^{1/(D-3)}$.
Physically, when the initial particles are wrapped within the horizon, a black hole forms.
It has been proven in four dimensions~\cite{Eardley:2002re} and in higher dimensions~\cite{Yoshino:2002tx} that a classical gravitational collision of two massless particles, whose gravitational fields are simulated by the Aichelburg-Sexl solution~\cite{Aichelburg:1970dh}, leads to a formation of a trapped surface, outside which there must be an event horizon~\cite{Hawking:1973uf}.
This argument has been generalized to the collision of two wave packets~\cite{Giddings:2004xy}.
The resultant black hole production cross section turns out to be geometrical 
\bea
\sigma_\text{BH}\sim R_{S}^{D-2}\sim(g_{s}^{2}M)^{D-2 \over D-3}.
\label{BH_cross_section}
\eea
At the correspondence point, the black hole production cross section~\eqref{BH_cross_section} becomes unity $\sigma_\text{BH}\sim 1$. Clearly, the tree level total cross section of string~\eqref{total_cross_section} and black hole do not coincide at the correspondence point $s\sim g_s^{-4}$.
We note that the classical treatment of the gravitational interactions is valid when the black hole mass is larger than the $D$~dimensional Planck scale $M\gtrsim M_{D}\sim g_{s}^{-2/(D-2)}$ and that $M_D$ is smaller than the correspondence scale $M_{C}\sim g_{s}^{-2}$.
Therefore the black hole production cross section can be trusted at $M\gtrsim M_{C}$ for $g_{s}<1$.
The meaning of the correspondence point is that the (classical) stringy corrections cannot be neglected for $M\lesssim M_{C}$ though the classical treatment of the gravity is still valid.
A cartoon is shown in Fig.~\ref{correspondence_fig} to help understanding.

As is emphasized in Introduction, it is important to consider the partial wave cross sections to 
see more detailed information.
For a black hole, the partial wave cross section is simply given by the differential cross section $d\sigma_\text{BH}/dJ$.
Let us consider a Kerr black hole with mass $M=\sqrt{s}$.
There are ${}_dC_2=d(d-1)/2$ components of angular momenta according to the choice of planes in $d$ spatial directions. The number of independent components are $d/2$ for $d$ even and $(d-1)/2$ for $d$ odd. 
Let us review the argument presented in~\cite{Ida:2002ez}.
Initial two light particles collide with an impact parameter $b$ in $D=d+1$ dimensions.
By choosing the scattering plane as 1--2 one, the initial system has only single non-zero angular momentum component $J=J_{12}=b\sqrt{s}/2$, with $\sqrt{s}$ being the center of mass energy and $b$ the impact parameter.
If angular momentum is conserved during the black hole formation process, it is sufficient to consider a black hole having only single non-zero angular momentum component. 
For such a higher dimensional rotating black hole, the horizon radius $r_h$ is determined by
\bea
r_h^{D-5}\left(r_h^2+{(D-2)^2J^2\over 4M^2}\right)={16\pi G M \over (D-2)\Omega_{D-2}} .
\eea
Utilizing the rescaled mass and angular momentum
\bea
\tilde{\mu} \equiv {16\pi G M \over (D-2)\Omega_{D-2}}, \qquad a_* \equiv {(D-2)J \over 2Mr_h} ,
\eea
the horizon radius can be written as
\begin{align}
	r_h	&=	\left({\tilde{\mu}\over1+a_*^2}\right)^{1\over D-3}.
		\label{horizon_radius}
  \end{align}
Note that the horizon radius $r_h$ is now given in terms of the mass $M=\sqrt{s}$ and the angular momentum $J=b\sqrt{s}/2$.
For a fixed $s$, one can show that $r_h$ is a decreasing function of~$b$.
Following from the hoop conjecture, a black hole would be produced when the impact parameter of the collision is less than the diameter of the black hole\footnote{However, this is not a coordinate invariant description. The diameter used here is provided by the Schwarzschild radius itself. See Appendix~\ref{rot_bh} for a related discussion on the radius.}
\begin{align}
b\leq2r_h(b). \label{hoop_conj}
\end{align}
For a given $M=\sqrt{s}$, the right hand side of Eq.~\eqref{hoop_conj} can be shown to be a decreasing function of $b$, and hence there is a maximum impact parameter that saturates the inequality~\eqref{hoop_conj}.
Noting that $b=2J/M=4r_ha_*/(D-2)$, the condition~\eqref{hoop_conj} leads to $a_*\leq(D-2)/2$, whose equality gives the maximum impact parameter
\begin{align}
b_\text{max}
	&=	2\left[{\tilde{\mu}\over1+a_{*\text{max}}^2}\right]^{{1\over D-3}}
\qquad
\text{with}\quad a_{*\text{max}}={D-2\over2}.
	\label{bmax}
\end{align}

It is amusing that $b_\text{max}$ exactly coincides with the naive Schwarzschild estimation $r_S=\tilde{\mu}^{1/(D-3)}$ in $D=4$ spacetime dimensions and that $b_\text{max}>r_S$ for $D\geq5$. The more dimensions we have, the bigger the increase of the $b_\text{max}$, the cross section.
This tendency agrees with the numerical analysis and the numerical values agrees within order ten percent accuracy in an appropriate number of (large) dimensions for string theory~\cite{Yoshino:2002tx,Yoshino:2005hi}.\footnote{In Ref.~\cite{Anchordoqui:2001cg}, quite a similar condition $b\leq r_h(b)$ was examined and it was concluded that the cross section would decrease from $r_S$ (corresponding to different $a_{*\text{max}}$).} Therefore we conclude that our assumption that the initial angular momentum and energy are almost all packed into the black hole is justified unless $b$ is not very close to $b_\text{max}$.\footnote{When $b\sim b_\text{max}$, the reduced mass, which gives the lower bound for the final black hole mass, is sizably reduced from the ``all-packed'' assumption~\cite{Yoshino:2002tx}.}

The effect of the angular momentum appear only through $a_*$ that is less than or of order one as explained above. 
Therefore one can drop the $a_*$-dependent factors when considering the total cross section 
up to a numerical factor of order one.\footnote{It is also true in the arugument of the entropy correspondence of rotating string/black hole. } 
Neglecting all such numerical coefficients, we get 
\bea
b_\text{max}\sim\left(G\sqrt{s}\right)^{1\over D-3} 
\eea
and the black hole cross section is 
\begin{align}
  \sigma_\text{BH}
    &\sim b_\text{max}^{D-2} \sim \left(G\sqrt{s} \right)^{{D-2\over D-3}}.
       \label{cross_section}
  \end{align}
With the above assumptions we also get the differential cross section of black hole with given angular momentum $J$ as 
\bea
  {d\sigma_\text{BH}\over dJ}
    \simeq \frac{J^{D-3}}{s^{D-2 \over 2}} \qquad
    \text{for} \quad J\lesssim J_\text{max}\sim \left(Gs^{D-2\over 2}\right)^{1\over D-3} .
	\label{dsigmadJ}
  \eea
where we have used $d\sigma_\text{BH}\simeq b^{D-3} db$.\footnote{It is interesting to notice that the maximal angular momentum $J_\text{max}$ and the entropy have the same form $(G s^{(D-2)/2})^{1/(D-3)}$. }

Finally, comparing Eqs.~\eqref{sigmaJ_app}, \eqref{sigmaJ_app_open} and \eqref{dsigmadJ} we have shown the correspondence of the partial wave cross sections between closed/open strings and black holes
\bea
\sigma_{J}^\text{string}(s) \leftrightarrow {d \sigma_\text{BH} \over dJ}
\eea
for $1 \ll J \leq \sqrt{s \ln s}$.

\section{Summary and discussion}
We have studied the production cross section of a highly excited string with a fixed angular momentum from a high energy collision of two light open or closed strings at the tree level.
We have also re-derived the partial wave unitarity condition at high energy.
We find that the cross sections exhibit, in addition to the softness in the large angular momentum/impact parameter space, geometric behaviors in the large but less than $\sqrt{\ln s}$ region of the impact parameter space.
The geometric behavior is characteristic to differential cross sections of black hole and thus we see a correspondence between string and black hole.

The total cross sections do not coincide at the correspondence point at the tree level analysis.
These cross sections of closed string and black hole share the growing behavior, both break the unitarity at high energy, while that of the open string stays constant.
On the other hand, we have presented that the behaviors of the partial wave cross sections are common between closed/open strings and black holes. 
Although our tree level amplitude for closed string cannot be trusted at the correspondence point because of the unitarity violation, the agreement of the geometric behaviors in the finite lower energy range is an incarnation of the correspondence principle.

Generically, the geometric behavior of the partial cross sections is a consequence of black disc total cross section.
Interestingly enough, high energy hadron collisions are also well described by the black disc approximation in which the radius of the disc is given by a range of strong interaction typically provided by a mass gap. 
The black disc provides Color Glass Condensation (CGC) in which gluon density saturates. The CGC is set as an initial condition to simulate the quark-gluon plasma generated in high energy hadron collisions, see e.g.~\cite{McLerran:2001sr,Iancu:2003xm}. 
Our string geometric cross sections might well be a model for the hadron and the CGC.

We are focusing on the $s$-channel single heavy string production, which is equivalent to the $t$-channel single graviton exchange.
If the correspondence holds at all, production process of the single string, dressed by the graviton cloud that accounts for the self interactions, would be more or less smoothly connected to that of the black hole.
However, Amati, Ciafaloni and Veneziano (ACV) showed long before~\cite{Amati:1987wq,Amati:1987uf} that a string feels as if it is propagating under Aichelburg-Sexl background~\cite{Aichelburg:1970dh} after summing over the eikonal graviton exchanges with the other string, see also Ref.~\cite{Veneziano:2004er} for more recent discussion.
Note that, as we already stated, a `collision' of two Aichelburg-Sexl solutions is proven to lead to a black hole production in four dimensions~\cite{Eardley:2002re} and in higher dimensions~\cite{Yoshino:2002tx}.
In this sense, the infinitely many eikonal $t$-channel graviton exchanges appear to provide the correspondence here.
Furthermore, the dominant contribution in the path integral is from the configuration where all the internal strings share equal amount of energy and hence there are no special one to be picked out~\cite{Giddings:2007bw}.
Therefore, it is somewhat puzzling how to reconcile the former correspondence picture with the latter ACV one. To repeat, the former involves a single $t$-channel massless string exchange (though with graviton clouds) while the latter is a sum over such exchanges, yet somehow to re-emerge a single-string-like behavior for the correspondence picture (a black hole being a long string at the threshold) to be recovered.

We have studied the production processes in which the initial and final objects in the collision are same, that is, $\text{open} + \text{open}$ to an open string and $\text{closed} + \text{closed}$ to a closed string. 
In order to understand differences and/or common features (universality), it would be interesting to study the production cross section of a rotating closed string from two light open strings on D$p$-brane, which is also more realistic to describe a scattering in the brane world scenario~\cite{Chialva:2005gt}, by performing similar analysis on the annulus amplitude with each pair of open strings attached on the inner and outer boundaries.

We hope to report investigations on the points mentioned above elsewhere. 

\section*{Acknowledgments}
We are grateful to Hikaru Kawai and Tsunehide Kuroki for useful discussions presented in~\cite{Kuroki:2007aj}, which provided basis for this work.
We thank Yoichiro Nambu for valuable comments.
We thank Roberto Emparan, Yoshitaka Hatta, Feng-Li Lin and Yuji Satoh for useful discussions.
T.M.\ also acknowledges the hospitality of Albert Einstein Institute, Niles Bohr Institute and theoretical high energy physics group at Pisa while this work was in progress.
We thank the RIKEN Theory Group where a large fraction of this work is done.
The work of K.O.\ is partially supported by Scientific Grant by Ministry of Education and Science, Nos.~19740171, 20244028, and 20025004.

\appendix
\section*{Appendix}
\section{Total cross section}
\label{sigma_from_residue}
In this Appendix we briefly review the tree level computation of the production cross section of the level $N$ resonance \cite{Dimopoulos:2001qe}.
We consider the Virasoro-shapiro amplitude \eqref{Virasoro_Shapiro} which can be written around an $s$-channel pole at $s=N$ as,
\begin{align}
	\cA(s,t)
		&=	{1\over s-N+i\epsilon}\Res_{s=N}\cA(s,t)
				+\text{(terms analytic at $s=N$)},
  \end{align}
where the infinitesimal $\epsilon$ gives
\begin{align}
	\im \cA(s,t)
		&=	-\pi\delta(s-N)\Res_{s=N}\cA(s,t)
			+\text{(terms analytic at $s=N$)}.
				\label{imaginary_A_peaked}
  \end{align}
Combining Eq.~\eqref{imaginary_A_peaked} with the optical theorem, the cross section around the $N$th resonance is given as
\begin{align}
	\sigma_N(s)
		&=
			-{\pi\over s}\delta(s-N)\Res_{s=N}\cA(s,0)
			+\text{(terms analytic at $s=N$)}.
			\label{tot_crossec_delta}
  \end{align}
Noting that the $s=N$ residue resides only in $\Res_{s=N}\Gamma(-1-s)={(-1)^N\over\Gamma(N+2)}$ in the amplitude~\eqref{Virasoro_Shapiro}, it is straightforward to compute
\begin{align}
	\Res_{s=N}\cA(s,t)
		&=	-2\pi g_s^2\left(\Gamma(N+t+3)\over\Gamma(N+2)\Gamma(t+2)\right)^2,
			\label{A_integrand}
  \end{align}
where we have also utilized $\Gamma(x)\Gamma(1-x)=\pi/\sin\pi x$.
Therefore, we obtain
\begin{align}
  \sigma_N(s)
	 &=	2\pi^2 g_s^2{(N+2)^2\over s}\delta(s-N)
	 	+\text{(terms analytic at $s=N$)}.
	 	\label{sigma_with_delta}
  \end{align}
Finally the total cross section becomes~\cite{Dimopoulos:2001qe}
\begin{align}
	\sigma_\text{tot}(s)
		&=	\sum_N\sigma_N(s)
		\simeq
			\sum_N 2\pi^2 g_s^2N\,\delta(s-N),
			\label{sigma_tot}
\end{align}
where the $N\gg1$ limit is taken in the last step.

The total cross section~\eqref{sigma_tot}, which has been given at the tree level, becomes zero and infinity off and at the resonances, respectively.
When we take the loop corrections into account, the $N$th resonance will have a finite decay width $\Gamma_N$ that corresponds to a finite $\epsilon_N\simeq2\sqrt{N}\Gamma_N$:
\begin{align}
	{1\over s-N+i\epsilon_N}
		&=	{s-N\over(s-N)^2+\epsilon_N^2}-i{\epsilon_N\over(s-N)^2+\epsilon_N^2}.
  \end{align}
When the decay width for the $N$th resonance is small enough, we can recover
\begin{align}
		\lim_{\epsilon_N\to0}{1\over s-N+i\epsilon_N}
			&\simeq	
				\cP{1\over s-N}-i\pi\delta(s-N).
				\label{resonance}
  \end{align}
Otherwise, the delta function is meant to give the correct value when integrated over the width of the peak.\footnote{Recall that $\int_{-\infty}^\infty ds{\epsilon\over(s-N)^2+\epsilon^2}=\pi$ regardless of the magnitude of $\epsilon$.}
In other words, at the large $N$ the spacing of the delta function is close and we may have 
\begin{align}
	\sigma_\text{tot}(s)
			\simeq 2\pi^2 g_s^2 s.
\end{align}
It should be kept in mind that, in reality, the decay width for the higher resonance $N\gg1$ can be large due to the exponentially growing number of the decay modes. 

The pole at the tree level amplitude~\eqref{Virasoro_Shapiro} does not have an imaginary part corresponding to the decay width of the resonance (other than the elastic one), which will be served by the higher loop corrections including many body final states. The sharp $s$-channel resonance will eventually be smeared out due to the exponential grow of the number of such decay channels. Note also that when one takes the Regge limit of a stringy amplitude, such a width is implicitly taken into account.

\section{Gegenbauer polynomial}
\label{Gegenbauer_formulas}
In this Appendix we collect some useful formulas for our computations.
For the gamma function:
\begin{align}
	\Gamma\!\left(\nu+{1\over2}\right)
		&=	{\sqrt{\pi}\over2^{2\nu-1}}{\Gamma(2\nu)\over\Gamma(\nu)}
		 =	{\sqrt{\pi}\over2^{2\nu-1}}\left(\nu\right)_\nu.
		 	\label{gamma_formula}
  \end{align}
The orthogonality condition for the Gegenbauer polynomial:
\begin{align}
	\int_{-1}^1dz\left(1-z^2\right)^{\nu-{1\over2}}C_J^\nu(z)\,C_{J'}^\nu(z)
		&=	N_J^\nu\,\delta_{JJ'},
			\label{Gegenbauer_orthogonality}
  \end{align}
where the normalization factor is given by
\begin{align}
	N_J^\nu
		&=	{\pi\Gamma(J+2\nu)\over2^{2\nu-1}(J+\nu)\,\Gamma(J+1)\,\Gamma(\nu)^2}
		 =	{\sqrt{\pi}\left(\nu\right)_{1/2}\over J+\nu}C_J^\nu(1).
		 	\label{normalization_factor}
  \end{align}
The angular integral formula for the Gegenbauer polynomial:
\begin{align}
	\int d\Omega_n C_J^\nu(\cos\theta_{fn})C_{J'}^\nu(\cos\theta_{in})
		&=	
			{(4\pi)^\nu N_J^\nu\over\left(\nu\right)_\nu C_J^\nu(1)}
			C_J^\nu(\cos\theta_{if})
			\delta_{JJ'}.
			\label{Gegenbauer_angular_integral}
  \end{align}
Another useful integral formula for $\re\nu>-1/2$:
\begin{align}
	\int_{-1}^1dz\left(1-z^2\right)^{\nu-{1\over2}}C_J^\nu(z)e^{\lambda z}
		&=	{2^\nu\sqrt{\pi}\Gamma\!\left(\nu+{1\over2}\right)C_J^\nu(1)I_{J+\nu}(\lambda)\over \lambda^\nu}.
			\label{Gegenbauer_integral_plus}
  \end{align}
For large $\lambda$, the modified Bessel function has the asymptotic form~\cite{Kuroki:2007aj}
\begin{align}
	I_n(\lambda)
		&\simeq	\frac{1}{\sqrt{2\pi\lambda}} 
				\exp\!\left(\lambda - {n^2\over2\lambda}\right).
				\label{I_limit}
  \end{align}

\section{Partial-wave unitarity bound}
\label{partial_wave_unitarity_bound}
We spell out the unitarity condition in our notation basically following Ref.~\cite{Soldate:1986mk}.
Set the $S$-matrix $S=1+iT$ and write generically $\braket{j|T|k}=(2\pi)^D\delta^D(P_j-P_k)\,\cA(k\to j)$. Let $\Ket{i}$ and $\Ket{f}$ be two-body states of the same particle contents so that $\braket{f|T|i}$ describes the corresponding elastic scattering. The unitarity condition $S^\dagger S=1$ implies $-i(T-T^\dagger)=T^\dagger T$, that is,
\begin{align}
	2\im\cA^\text{el}(i\to f)
		&=	\sum_n(2\pi)^D\delta^D(P_n-P_i)\,\left[\cA(f\to n)\right]^*\,\cA(i\to n),
			\label{unitarity_relation}
  \end{align}
where the summation over $n$ includes momentum integrals.
We can separate the sum into elastic and other parts $\sum_n=\sum_n^\text{el}+\sum_n^\text{others}$. In the first elastic sum, $\ket{n}$ has the same particle contents as $\ket{i}$ and $\ket{j}$, while the second sum includes inelastic scattering, many body final states, etc.

From now on, we work in the center of mass frame unless otherwise stated. 
The phase space integral reads
\begin{align}
	\sum_n^\text{elastic}(2\pi)^D\delta^D(P_n-P_i)
		&=	\int{d^d\mathbf{p}_{n1}\over(2\pi)^d2E_{n1}}
			\int{d^d\mathbf{p}_{n2}\over(2\pi)^d2E_{n2}}
			(2\pi)^D\delta^D(p_{n1}+p_{n2}-P_i)\nn
		&=	{|\mathbf{p}_{n1}|^{D-3}\over(2\pi)^{D-2}4P_i^0}\int d\Omega_{n1}
		\to	{s^{\nu-{1\over2}}\over2(4\pi)^{2\nu+1}}\int d\Omega_{n1},
			\label{phase_space_integral}
  \end{align} 
where we used $P_i^0=\sqrt{s}$ and also $|\mathbf{p}_{n1}|\to\sqrt{s}/2$ for $s$ much larger than the corresponding mass squared.
We expand Eq.~\eqref{unitarity_relation} into partial waves by Eq.~\eqref{partial_wave_expansion}
\begin{align}
	& 2\sum_{J=0}^\infty\im\cA_J^\text{el}(i\to f){C_J^{\nu}(\cos\theta_{if})\over C_J^\nu(1)}\nn
		&\quad=	\sum_n(2\pi)^D\delta^D(P_i-P_n)\sum_{J=0}^\infty\left[\cA_J(f\to n)\right]^*\frac{C_J^\nu(\cos\theta_{fn})}{C_J^\nu(1)}\sum_{J'=0}^\infty\cA_{J'}(i \to n)\frac{C_{J'}^\nu(\cos\theta_{in})}{C_{J'}^\nu(1)}\nn
		&\quad=	{s^{\nu-{1\over2}}\over2(4\pi)^{2\nu+1}}
				\sum_{J,J'=0}^\infty
				\left[\cA_J^\text{el}(f\to n)\right]^*\cA_{J'}^\text{el}(i\to n)
				\int d\Omega_{n}
				{C_J^{\nu}(\cos\theta_{fn})C_{J'}^{\nu}(\cos\theta_{in})\over
					C_J^\nu(1)C_{J'}^\nu(1)}+\text{others}\nn
		&\quad=	{s^{\nu-{1\over2}}\over2(4\pi)^{\nu+1}}
				{\Gamma(\nu)\over\Gamma(2\nu)}
				\sum_{J=0}^\infty
				\left[\cA_J^\text{el}(f\to n)\right]^*\cA_J^\text{el}(i\to n)				
				{N_J^\nu\over{C_J^\nu(1)}^2}{C_J^\nu(\cos\theta_{if})\over C_J^\nu(1)}
				+\text{others},
				\label{partial_wave_unitarity}
  \end{align}
where ``others'' denotes $\sum_n^\text{others}\left[\cA(f\to n)\right]^*\cA(i\to n)\,(2\pi)^D\delta^D(p_n-p_i)$ and we have utilized Eq.~\eqref{phase_space_integral} and \eqref{Gegenbauer_angular_integral}.

In the forward scattering limit $\ket{f}\to\ket{i}$, each term in the sum $\sum_n$ in Eq.~\eqref{partial_wave_unitarity} goes to $\left|\cA(i\to n)\right|^2$ and becomes positive. Noting that the elastic matrix elements depend only on $s$ and $J$, namely $\cA^\text{el}_J(i\to n)=\cA_J^\text{el}(s)$ for any $n$, we get a sufficient condition for each $J$ in order to satisfy the unitarity of the S-matrix:
\begin{align}
	2\im \cA_J^\text{el}(s)
		&\geq
			{s^{\nu-{1\over2}}\over2(4\pi)^{\nu+1}}
			{\Gamma(\nu)\over\Gamma(2\nu)}
			{N_J^\nu\over{C_J^\nu(1)}^2}
			\left|\cA_J^\text{el}(s)\right|^2.
			\label{unitarity_sufficient_condition}
  \end{align}
Defining 
\begin{align}
	a_J(s)	\equiv	{s^{\nu-{1\over2}}\over4(4\pi)^{\nu+1}}
					{\Gamma(\nu)\over\Gamma(2\nu)}
					{N_J^\nu\over{C_J^\nu(1)}^2}
					\cA_J^\text{el}(s),
					\label{normalized_amplitude}
  \end{align}
the unitarity condition~\eqref{unitarity_sufficient_condition} reads
\begin{align}
	\im a_J(s) \geq |a_J(s)|^2.
	 \label{unitarity_condition_for_normalized_amplitude}
\end{align}
As an immediate corollary
\begin{equation}
\sqrt{\left(\re a_J\right)^2+\left(\im a_J-{1\over2}\right)^2}\leq{1\over2}
\end{equation} 
or
\begin{align}
	|a_J(s)|\leq1.
		\label{final_unitarity_bound}
  \end{align}

\section{Various radii and ultra-spinning black disk formation}
\label{rot_bh}
In this Section we comment on some subtleties on the choice of the horizon radius which is to be used in the naive estimation of the cross section. 
Generically one should not take the numerical coincidence of the total cross section between the naive one computed from Eq.~\eqref{bmax} and the exact lower bound by Refs.~\cite{Yoshino:2002tx,Yoshino:2005hi} too literally, because the definition of the radius~\eqref{horizon_radius} is not coordinate invariant.\footnote{
We thank Roberto Emparan for the discussion on which this section is based.} 
Our metric for the relevant Myers-Perry black hole with a single angular momentum is given by
\begin{align}
	ds^2
		&=	-dt^2
			+{\mu\over r^{D-5}\rho^2}\left(dt+a\sin^2\theta d\phi\right)^2
			+{\rho^2\over\Delta}dr^2
			+\rho^2d\theta^2
			+(r^2+a^2)\sin^2\theta d\phi^2
			+r^2\cos^2\theta d\Omega^2_{D-4},
			\label{MP_metric}
  \end{align}
where
\begin{align}
	\rho^2(r,\theta)
		&=	r^2+a^2\cos^2\theta, &
	\Delta(r)
		&=	r^2+a^2-{\mu\over r^{D-5}}.
  \end{align}

We can define the following proper radii for a rotating black hole in $D$~dimensions
\begin{align}
	r_\text{tot}
		\equiv
			\left(A_\text{tot}/\Omega_{D-2}\right)^{1/(D-2)}
		&=	\left(1+a_*^2\right)^{1/(D-2)}r_h,\\
	r_\parallel
		\equiv
			\left(A_\parallel/4\pi\right)^{1/2}
		&=	\left(1+a_*^2\right)^{1/2}r_h,	\label{r_para}\\
	r_\text{eq}	
		\equiv
			\ell_\text{eq}/2\pi
		&=	\left(1+a_*^2\right)r_h,
  \end{align}
where
	$A_\text{tot}$ is the total horizon area,
	$A_\parallel$ is the two dimensional area along $\theta$ and $\phi$ directions,\footnote{Note that $A_\parallel$ is not the area of the section of the horizon that lies on the brane.} and 
	$\ell_\text{eq}$ is the length of the equator.\footnote{More explicitly, the length $\ell_\text{eq}$ is the integral of $\sqrt{g_{\phi\phi}}$ over $\phi=0$ to $2\pi$ with fixed $r=r_h$ and $\theta=\pi/2$ in the metric~\eqref{MP_metric}, etc.}
For a very large angular momentum $a_*\gg 1$, the resultant ultra-spinning black hole takes the shape of a thin pancake whose thickness being of order $r_h$, defined in Eq.~\eqref{horizon_radius}, while the pancake radius being of order $a_*r_h\gg r_h$~\cite{Emparan:2003sy}. In this regard, the pancake radius should correspond to~\eqref{r_para}.

The estimation~\eqref{bmax} that uses the shortest definition of the horizon size~\eqref{horizon_radius} provides the most conservative lower bound on $b_\text{max}$ within this naive estimation framework. 
If we require the impact parameter of the collision to be smaller than the diameter with respect to this pancake radius
\begin{align}
	b<2r_\text{eq}(b)
		\label{ultra_spinning}
  \end{align}
instead of~\eqref{hoop_conj}, the condition~\eqref{ultra_spinning} can be satisfied by an arbitrary large $b$ (and hence $a_*$) without leading to an upper bound on $b$ and therefore the cross section diverges naively.
In such an extreme, the upper bound on $b$ and hence on $a_*$ would be put by requiring the pancake thickness to be longer than the Planck length $r_h\gtrsim M_D^{-1}$ for a fixed $M$ ($\sim\sqrt{s}$), instead of the classical naive consideration~\eqref{ultra_spinning}.
We note that such an ultra-spinning black hole with $a_*\gg 1$ will suffer from classical gravitational instabilities~\cite{Emparan:2003sy}, which might lead to a formation of a black ring~\cite{Ida:2002ez} that will also suffer from the black string instabilities.\footnote{
In Ref.~\cite{Ida:2002ez} and its series, terminology of ``large angular momentum'' is used to indicate that the Hawking radiation is greatly distorted from the Schwarzschild one, which is generically the case even for $a_*\lesssim 1$. 
For example, the maximum angular parameter from the conservative estimate~\eqref{hoop_conj} takes values $a_{*\text{max}}=1.5$ to 4 for $D=5$ to 10.} 
In this paper, we constrain ourselves within more conservative range $a\leq a_{*\text{max}}=(D-2)/2$, leaving a stringy consideration of the possible case $a_*\gg1$ for future research.


\begin{thebibliography}{99}



\bibitem{Susskind:1993ws}
  L.~Susskind,
  ``Some speculations about black hole entropy in string theory,''
  arXiv:hep-th/9309145.


\bibitem{Horowitz:1996nw}
  G.~T.~Horowitz and J.~Polchinski,
  ``A correspondence principle for black holes and strings,''
  Phys.\ Rev.\  D {\bf 55} (1997) 6189
  [arXiv:hep-th/9612146].



\bibitem{Horowitz:1997jc}
  G.~T.~Horowitz and J.~Polchinski,
  ``Self gravitating fundamental strings,''
  Phys.\ Rev.\  D {\bf 57} (1998) 2557
  [arXiv:hep-th/9707170].
  
\bibitem{Damour:1999aw}
  T.~Damour and G.~Veneziano,
  ``Self-gravitating fundamental strings and black holes,''
  Nucl.\ Phys.\  B {\bf 568} (2000) 93
  [arXiv:hep-th/9907030].




\bibitem{Antoniadis:1998ig}
  I.~Antoniadis, N.~Arkani-Hamed, S.~Dimopoulos and G.~R.~Dvali,
  ``New dimensions at a millimeter to a Fermi and superstrings at a TeV,''
  Phys.\ Lett.\  B {\bf 436} (1998) 257
  [arXiv:hep-ph/9804398].

\bibitem{Randall:1999ee}
  L.~Randall and R.~Sundrum,
  ``A large mass hierarchy from a small extra dimension,''
  Phys.\ Rev.\ Lett.\  {\bf 83} (1999) 3370
  [arXiv:hep-ph/9905221].

\bibitem{Giddings:2001bu}
  S.~B.~Giddings and S.~D.~Thomas,
  ``High energy colliders as black hole factories: The end of short  distance
  physics,''
  Phys.\ Rev.\  D {\bf 65} (2002) 056010
  [arXiv:hep-ph/0106219].

\bibitem{Dimopoulos:2001hw}
  S.~Dimopoulos and G.~L.~Landsberg,
  ``Black holes at the LHC,''
  Phys.\ Rev.\ Lett.\  {\bf 87} (2001) 161602
  [arXiv:hep-ph/0106295].


\bibitem{Oda:2001uw}
  K.~Oda and N.~Okada,
  ``Alternative signature of TeV strings,''
  Phys.\ Rev.\  D {\bf 66}, 095005 (2002)
  [arXiv:hep-ph/0111298].
  
  
    \bibitem{Eardley:2002re}
  D.~M.~Eardley and S.~B.~Giddings,
  ``Classical black hole production in high-energy collisions,''
  Phys.\ Rev.\  D {\bf 66} (2002) 044011
  [arXiv:gr-qc/0201034].
  
\bibitem{Yoshino:2002tx}
  H.~Yoshino and Y.~Nambu,
  ``Black hole formation in the grazing collision of high-energy particles,''
  Phys.\ Rev.\  D {\bf 67} (2003) 024009
  [arXiv:gr-qc/0209003].
  
  \bibitem{Yoshino:2005hi}
  H.~Yoshino and V.~S.~Rychkov,
  ``Improved analysis of black hole formation in high-energy particle
  collisions,''
  Phys.\ Rev.\  D {\bf 71} (2005) 104028
  [arXiv:hep-th/0503171].
  
\bibitem{'t Hooft:1987rb}
  G.~'t Hooft,
  ``Graviton Dominance in Ultrahigh-Energy Scattering,''
  Phys.\ Lett.\  B {\bf 198} (1987) 61.

\bibitem{Dimopoulos:2001qe}
  S.~Dimopoulos and R.~Emparan,
  ``String balls at the LHC and beyond,''
  Phys.\ Lett.\  B {\bf 526} (2002) 393
  [arXiv:hep-ph/0108060].

\bibitem{Nambu-san}
	Y.~Nambu, private communication.


\bibitem{Ida:2002ez}
  D.~Ida, K.~Oda and S.~C.~Park,
  ``Rotating black holes at future colliders: Greybody factors for brane
  fields,''
  Phys.\ Rev.\  D {\bf 67}, 064025 (2003)
  [Erratum-ibid.\  D {\bf 69}, 049901 (2004)]
  [arXiv:hep-th/0212108].


\bibitem{Veneziano:2004er}
  G.~Veneziano,
  ``String-theoretic unitary S-matrix at the threshold of black-hole
  production,''
  JHEP {\bf 0411}, 001 (2004)
  [arXiv:hep-th/0410166].
\bibitem{Veneziano:2006wn}
  G.~Veneziano,
  ``Towards a unitary S-matrix description of black-hole formation and decay in
  string theory,''
  AIP Conf.\ Proc.\  {\bf 861}, 39 (2006).
  
\bibitem{Yoneya:2000bt}
  T.~Yoneya,
  ``String theory and space-time uncertainty principle,''
  Prog.\ Theor.\ Phys.\  {\bf 103} (2000) 1081
  [arXiv:hep-th/0004074].
  
  
\bibitem{Gross:1987kza}
  D.~J.~Gross and P.~F.~Mende,
  ``The High-Energy Behavior of String Scattering Amplitudes,''
  Phys.\ Lett.\  B {\bf 197} (1987) 129.
  
\bibitem{Gross:1987ar}
  D.~J.~Gross and P.~F.~Mende,
  ``String Theory Beyond the Planck Scale,''
  Nucl.\ Phys.\  B {\bf 303} (1988) 407.
  
\bibitem{Kuroki:2007aj}
  T.~Kuroki and T.~Matsuo,
  ``Production cross section of rotating string,''
  Nucl.\ Phys.\  B {\bf 798} (2008) 291
  [arXiv:0712.4062 [hep-th]].
  

\bibitem{Soldate:1986mk}
  M.~Soldate,
  ``Partial Wave Unitarity and Closed String Amplitudes,''
  Phys.\ Lett.\  B {\bf 186} (1987) 321.

\bibitem{Muzinich:1987in}
  I.~J.~Muzinich and M.~Soldate,
  ``High-Energy Unitarity of Gravitation and Strings,''
  Phys.\ Rev.\  D {\bf 37} (1988) 359.
  
\bibitem{Nambu:1969sb}
  Y.~Nambu and P.~Frampton,
  ``Asymptotic behavior of partial widths in the Veneziano model of scattering amplitudes,''
  Published in Quanta: Essays in Theoretical Physics Dedicated to Gregor Wentzel. Chicago, Ill., Univ. of Chicago Press, 1970. pp. 403-414.
  
\bibitem{Lin:2007gi}
  F.~L.~Lin, T.~Matsuo and D.~Tomino,
  ``Hagedorn Strings and Correspondence Principle in AdS(3),''
  JHEP {\bf 0709} (2007) 042
  [arXiv:0705.4514 [hep-th]].
  
\bibitem{Aichelburg:1970dh}
  P.~C.~Aichelburg and R.~U.~Sexl,
  ``On the Gravitational field of a massless particle,''
  Gen.\ Rel.\ Grav.\  {\bf 2} (1971) 303.
  
\bibitem{Hawking:1973uf}
  S.~W.~Hawking and G.~F.~R.~Ellis,
  ``The Large scale structure of space-time,''
{\it  Cambridge University Press, Cambridge, 1973}
  
    \bibitem{Giddings:2004xy}
  S.~B.~Giddings and V.~S.~Rychkov,
  ``Black holes from colliding wavepackets,''
  Phys.\ Rev.\  D {\bf 70} (2004) 104026
  [arXiv:hep-th/0409131].


  \bibitem{Anchordoqui:2001cg}
  L.~A.~Anchordoqui, J.~L.~Feng, H.~Goldberg and A.~D.~Shapere,
  ``Black holes from cosmic rays: Probes of extra dimensions and new limits  on
  TeV-scale gravity,''
  Phys.\ Rev.\  D {\bf 65}, 124027 (2002)
  [arXiv:hep-ph/0112247].


\bibitem{Emparan:2003sy}
  R.~Emparan and R.~C.~Myers,
  ``Instability of ultra-spinning black holes,''
  JHEP {\bf 0309} (2003) 025
  [arXiv:hep-th/0308056].
  
  

\bibitem{McLerran:2001sr}
L.~D.~McLerran,
``The color glass condensate and small x physics: 4 lectures,''
Lect.\ Notes Phys.\  {\bf 583} (2002) 291
[arXiv:hep-ph/0104285].

\bibitem{Iancu:2003xm}
  E.~Iancu and R.~Venugopalan,
  ``The color glass condensate and high energy scattering in QCD,''
  arXiv:hep-ph/0303204.
  
  
\bibitem{Amati:1987wq}
  D.~Amati, M.~Ciafaloni and G.~Veneziano,
  ``Superstring Collisions at Planckian Energies,''
  Phys.\ Lett.\  B {\bf 197} (1987) 81.
\bibitem{Amati:1987uf}
  D.~Amati, M.~Ciafaloni and G.~Veneziano,
  ``Classical and Quantum Gravity Effects from Planckian Energy Superstring
  Collisions,''
  Int.\ J.\ Mod.\ Phys.\  A {\bf 3} (1988) 1615.


\bibitem{Giddings:2007bw}
  S.~B.~Giddings, D.~J.~Gross and A.~Maharana,
  ``Gravitational effects in ultrahigh-energy string scattering,''
  Phys.\ Rev.\  D {\bf 77} (2008) 046001
  [arXiv:0705.1816 [hep-th]].

\bibitem{Chialva:2005gt}
  D.~Chialva, R.~Iengo and J.~G.~Russo,
  ``Cross sections for production of closed superstrings at high energy
  colliders in brane world models,''
  Phys.\ Rev.\  D {\bf 71} (2005) 106009
  [arXiv:hep-ph/0503125].


  

\end{thebibliography}
\end{document}